\documentclass[11pt]{article}

\usepackage{aaspp4}
\usepackage{psfig}
\usepackage{graphics}
\newcommand{\et}{PKS~1830$-$211}
\newcommand{\etal}{{\em et al.}}

\newcommand{\delay}{26}
\newcommand{\delayEplus}{5}
\newcommand{\delayEminus}{4}
\newcommand{\delayE}{$-\delay^{+\delayEplus}_{-\delayEminus}$}
\newcommand{\delayEx}{$\delay^{+\delayEminus}_{-\delayEplus}$}

\newcommand{\magnif}{1.52}

\newcommand{\magnifE}{$\magnif \pm 0.05$}

\newcommand{\foff}{-0.62}

\newcommand{\foffE}{$\foff \pm 0.05$}

\slugcomment{}

\lefthead{J.E.J.~Lovell {\em et al.}}

\begin{document}
\title{The Time Delay in the Gravitational Lens  PKS~1830$-$211}

\author{J.E.J.~Lovell\altaffilmark{1,2},
  D.L.~Jauncey\altaffilmark{3},  J.E.~Reynolds\altaffilmark{3}, 
  M.H.~Wieringa\altaffilmark{3}, E.A.~King\altaffilmark{3,4},
  A.K.~Tzioumis\altaffilmark{3}, P.M.~McCulloch\altaffilmark{1},
  P.G.~Edwards\altaffilmark{5}\\}

\authoraddr{J.E.J. Lovell, Institute of Space and Astronautical
  Science, 3-1-1 Yoshinodai, Sagamihara 229-8510, Kanagawa, Japan}
\authoremail{jlovell@vsop.isas.ac.jp}

\altaffiltext{1}{School of Mathematics and Physics, University of Tasmania, 
  Hobart, Tasmania 7001, Australia}
\altaffiltext{2}{Now at Institute of Space and Astronautical Science,
 Sagamihara, Kanagawa 229-8510, Japan}  
\altaffiltext{3}{Australia Telescope National Facility, CSIRO, Epping,
New South Wales 2121, Australia}
\altaffiltext{4}{Present address CSIRO Earth Observation Centre, Canberra,
                  ACT, 2601, Australia}
\altaffiltext{5}{Institute of Space and Astronautical Science,
 Sagamihara, Kanagawa 229-8510, Japan}

\begin{abstract}
We have measured a time delay of \delayEx\ days and a magnification
ratio of \magnifE\ in the strong radio gravitational lens \et. The
observations were made over the 18 month period from 1997 January to
1998 July with the Australia Telescope Compact Array at 8.6~GHz, and
have shown that the source started a large flux density
outburst around 1997 June.
\end{abstract}

\keywords{galaxies: individual (PKS~1830$-$211) --- Gravitational lensing}

\newpage

\section{Introduction}
A precise measurement of the time delay between the images in a
gravitational lens is essential if it is to fulfill its potential as a
tool for estimating the Hubble constant.  The flat-spectrum radio
source \et\ (\cite{rao88}) was found to be an Einstein
ring/gravitational lens, comprised of two compact, flat-spectrum
components located on opposite sides of a 1 arcsec diameter ring
(\cite{jau91}). It is by far the strongest radio lens yet found, being
$\sim10$~Jy at 4.8~GHz. Observations of molecular absorption at
centimeter and millimeter wavelengths have revealed two intervening
galaxies at $z=0.19$ (\cite{lov96a}) and $z=0.89$ (\cite{wik96}). VLBI
observations of the two compact lensed components have revealed
striking structural differences on milliarcsecond scales
(\cite{gar96}) suggesting that both galaxies may be involved in the
lensing (\cite{lov96a})\@. \et\ varies dramatically at radio
wavelengths (\cite{lov96b}) and VLBI observations show that the
variations are confined to the two compact components (\cite{kin94}),
making it an excellent candidate for relative time delay measurements.

This paper describes our measurements with the $6 \times 22$~m
Australia Telescope Compact Array (ATCA) to determine the lensing time
delay in \et\ through a correlation of the flux density light curves
of its compact components. Section 2 describes the development of our
observing strategy and data reduction, section 3 describes the
analysis of our data to obtain a time delay and magnification
ratio. In section 4 we describe our error analysis and in section 5 we
discuss the implications of our measurements on modeling the \et\
lensing system.

\section{Observations and Data Reduction}
\label{sec:obs}
The presence of a single peak in the 8.4~GHz total flux density light
curve in early 1992 (\cite{lov96b}) indicates that the lensing time
delay is less than a few months (a time delay that was much longer
would result in a broader or double-peaked light curve)\@.
Consequently, in order to sufficiently time resolve these variations
in the compact components it was necessary to measure the flux
densities of the two components separately on timescales shorter than
both the total flux density variations and the lensing time delay,
previously estimated to be $44 \pm 9$~days (van Ommen \etal\ 1995). We
commenced our ATCA monitoring observations in 1995 August, making
observations at approximately monthly intervals.  The two compact and
variable components can be measured separately, as explained below, on
the 6 km East-West array of the ATCA at 8.6~GHz if observations are
restricted to between 15 and 19 hours LST when the ATCA provides
sub-arcsecond spatial resolution along the position angle of the
compact components. The strong, compact, flat-spectrum source
PKS~1921--293 was observed in each session to provide the relative
gain calibration of the individual antennas, while overall flux
density calibration was determined through observations of the ATCA
primary flux density calibrator PKS~1934--638. The total flux density
variations of \et\ will be presented in a future paper (Lovell \etal\
1998, in preparation)\@. During the period 1996 July 26 through 1996
August 8, a more closely spaced series of observations was undertaken
in order to gain an understanding of the shorter-term variability.
Analysis of this 1996 July-August data showed more rapid variability
than previously thought, so from 1997 January, one observation every
three to six days was made to better sample the light curves. In this
paper we restrict our analysis to the data from 1996 July onwards,
although the data from the period 1996 December 20 through 1997
January 10 were also excluded as the source was too close to the sun.

As the two compact components of \et\ are not fully resolved from the
ring at the $\sim 0.9$~arcsecond resolution of the ATCA at 8.6~GHz, we
have adopted a model fitting approach to determine the flux densities
of the individual components.  We chose a simple model consisting of
two circular Gaussian components, each of which also contains a
contribution from the non-varying ring. The component separation was
set initially to 0.95 arcseconds at a position angle of 46 degrees, as
determined from VLBI observations (\cite{kin94}). These values were
allowed to change slightly during model fitting, but were found to
deviate by no more than 3 degrees in position angle and 12 mas in
separation. All data were edited and calibrated using AIPS before
being exported to Difmap (\cite{she97}) for model fitting.
 
To verify that this model successfully describes the source, we
analyzed an edited section from a full ATCA synthesis observation
taken on 1994 December 19 when \et\ had a total flux density of
7.2~Jy.  We found that the individual fitted Gaussian component flux
densities differed by no more than 70~mJy from those measured from the
full synthesis data.  As a further check, simulated ATCA data sets
were generated and analyzed using the above procedures. For all of
these the modeled component flux density changes were accurate to
within 1\%, and the total flux density in the model accounted for
99.8\% of the total simulated flux density. This clearly demonstrates
the robustness and reliability of our model fitting procedures in
determining small changes in each of the compact components of \et.
 
Component flux density errors were estimated at each epoch using the
method described by Tzioumis {\em et al.} (1989) and were found to be
typically $\sim50$~mJy. The NE and SW component flux density light
curves obtained from the fitted $8.6$~GHz ATCA data since 1996 January
are shown in Figure~1(a). We have detected clear changes in the flux
densities of the two components, with a roughly linear increase seen
to begin sometime between MJD 50530 and MJD 50670\@. There is a clear
difference in gradient between the two light curves, which is a result
of different image magnifications. The figure also shows some
significant differences in the light curves, the most prominent being
a distinct `bump' of 400~mJy in the NE component light curve near MJD
50700 that is seen to occur significantly later in the SW light curve,
indicating a time delay of 20 to 30 days.  In the following section,
we discuss a detailed analysis of the data to obtain a quantitative
estimate of the time delay and magnification ratio and their
uncertainties.

\section{Analysis}
To estimate the time delay it is important to consider how the
variability of the lensed source is detected by the observer. In the
following analysis, the subscripts 1 and 2 refer to the NE and SW
components of \et\ respectively. We describe the time variation of the NE
and SW compact components of the lensed image as $S_{1}(t)$ and
$S_{2}(t)$ respectively.  The SW component light curve is expected to
be identical to the NE light curve but magnified by a different amount
and shifted in time, thus:
\begin{equation}
\label{equ:two}
  S_{2}(t) = \frac{1}{\mu}S_{1}(t + \Delta \tau) 
\end{equation} 
where $\Delta \tau$ is the time delay and $\mu$ is the relative
magnification ratio. Also lensed is an extended component that forms
the Einstein ring which contributes a constant flux density
$S_{\mbox{\footnotesize ring}}$\@. The total observed flux density can
therefore be described as
\begin{equation} S_{m}(t) = S_{1}(t) + S_{2}(t) +
  S_{\mbox{\footnotesize ring}}.
\end{equation}
  
The Einstein ring contributes a large proportion, up to 50\% of the
total flux density at cm wavelengths (\cite{jau91})\@.  Therefore the
effect of the ring flux density must be considered before attempting
to determine a lensing time delay.  In our ATCA observations, neither
$S_{1}(t)$ nor $S_{2}(t)$ are measured directly, instead the measured
quantities are
\begin{equation} S_{m1}(t) = S_{1}(t) + S_{c1} \end{equation} and
\begin{equation} S_{m2}(t) = S_{2}(t) + S_{c2} = \frac{1}{\mu}S_{1}(t
+ \Delta \tau) + S_{c2} \end{equation} where the quantities $S_{c1,2}$
are constant but unknown and $S_{\mbox{\footnotesize ring}} = S_{c1} +
S_{c2}$\@.  It is unlikely that $S_{c1}$ and $S_{c2}$ are
equal. Therefore it is also necessary to solve for $d$, the difference
between $S_{c1}$ and $S_{c2}$.  The two light curves may now be
compared using trial values of $\mu$, $\Delta \tau$ and $d$ so that a
solution for the relative magnification ratio and the time delay can
be obtained.

\subsection{Dispersion Analysis}
A correlation of the two light curves was undertaken using the
dispersion analysis method introduced by Pelt \etal\ (1994, 1996) to
analyze the component light curves of the gravitational lens
0957$+$561\@.  The dispersion method has been shown to be successful
for 0957$+$561, which has a relatively large time delay of $417 \pm
3$~days (\cite{kun97})\@.  We chose this method because it avoids any
interpolation between data points. Methods that are based on
interpolation can lead to erroneous results as they put equal weight
on assumed and measured data.

Following Pelt {\em et al.}, for every test value of $\mu$, $\Delta
\tau$ and $d$, two light-curve datasets $a_i$ and $b_i$ ($i =
1,\ldots,N$ where $N$ is the number of observations) are
obtained. Together they form a combined light-curve where $a_i$
contains the NE variable component data and $b_i$ contains the SW variable
component data modified in flux density and time by the test values for
$\mu$, $\Delta \tau$ and $d$\@. The dispersion, $D^2$, of this
combined light curve is calculated from the weighted sum of squared
differences between nearby $a_{i},b_{i}$ pairs over the entire curve:

\begin{equation}
D^2(\Delta \tau,\mu) = \frac{\Sigma_{ij} W_{ij} V'_{ij} (a_i -
  b_j)^2}{2\Sigma_{ij} W_{ij} V'_{ij}} 
\label{equ:dsquared}
\end{equation}
where
\begin{equation}  
\label{equ:weight_1}
V'_{ij}=
\left\{ \begin{array}{ll}
 1        & \mbox{if $\left| t_i - t_j \right|  \le \delta$},\\ \\
 \left( 1 + {\left( \frac{\delta - \left|t_i - t_j \right|}
                            {\delta/2}\right)}^2        \right)^{-1}            & \mbox{if $\left| t_i - t_j \right| >  \delta$}.
\end{array}
\right.
\end{equation}
\begin{equation}
W_{ij} = \frac{W_i W_j}{W_i + W_j}
\end{equation}

and $W_{i} = 1/{e_{i}^{2}}$ where $e_{i}$ is the standard error in
data point $i$\@. The weighting factor $V'_{ij}$ ensures that data
pairs are only given significant weighting if they are less than
$\sim\delta$ days apart and is modified from the $V_{ij}$ of Pelt
\etal\ to avoid strong weighting on data points with near-zero
separation. These points add a bias towards a zero delay for
relatively short time delays, as we expect to find in \et.  The values
of $\Delta\tau$, $\mu$ and $d$ for which $D^2$ is a minimum are thus
the best estimate of the time delay, magnification ratio and constant
flux density offset difference.

Before a correlation of the light curves can be made, an estimate of
$\delta$, effectively the longest timescale over which the source is
believed not to show significant variability, must be determined. An
inspection of the light curves in Figure~1, shows significant changes
in flux density on timescales of the order of 10~days. We note that
$V'$ drops quite slowly, reaching 0.5 at a point separation of
$1.5\delta$. Therefore, we chose a value of $\delta=5$, slightly less
than the estimated timescale, so that points with separations greater
than $\sim 8$~days did not unduly influence the solution.

A single, unambiguous solution was found for $\delta=5$~days at
$\Delta \tau = -23$~days ({\em i.e.} NE component leading), $\mu =
\magnif$ and $d = \foff$~Jy. We also investigated how the solution
behaved for a range of $\delta$\@ and found that for $\delta <
7$~days, $\Delta \tau$ stayed within $\pm 4$~days, $\mu$ stayed within
$\pm0.02$ and $d$ stayed within $\pm 0.06$~Jy of the $\delta = 5$~days
solution. For $\delta \geq 7$~days, up to $\delta = 14$~days (the
largest value we trialed), the solutions for $\Delta \tau$ became
shorter, ranging from $-12$ to $-17$ days, but solutions for $\mu$ and
$d$ changed little. A visual inspection of the combined light curves
with these solutions applied clearly shows a poorer correlation than
seen for $\delta < 7$\@.

 It is likely that this sudden change in the delay solution is due to
an effective smoothing of the features that help register the light
curves in delay. However, $\mu$ and $d$ are affected to a lesser
degree as the overall smooth, rising light curves constrain these
parameters well, and the short timescale changes have little influence. 

\section{Error Analysis}
We carried out Monte Carlo simulations to estimate the errors in our
derivation of the time delay and magnification ratio. Five hundred
light curve pairs were created with data points at the same epochs of
the actual observations. The flux density of each point was calculated
using a random number generator but constrained such that each
simulated point was within the uncertainty of the original
measurement. An additional constraint forced the sum of a pair of
same-epoch measurements to be equal to the total measured flux density
within the uncertainty of that observation.

Each of the 500 light curve pairs were then correlated to obtain
estimates for $\Delta \tau$, $\mu$ and $d$ when $\delta=5$~days.  We
found that the correlated values of $\mu$ and $d$ were distributed in a
Gaussian-like manner about the values obtained from our
observations. We have used these distributions to estimate errors for
$\mu$ and $d$ by measuring the width of the distributions at half
their peak. We thus estimate that the magnification ratio, $\mu$, is
\magnifE\ and the flux density offset, $d$, is \foffE~Jy.
The distribution of $\Delta \tau$ from the simulations however is not
distributed in a Gaussian-like fashion. Instead, a broad distribution
of solutions between --12 and --30~days was obtained with peaks near
--15, --23 and --28~days.

Eighteen years of monitoring the gravitational lens 0957$+$561 have
demonstrated that non-varying or smooth light curves contributes
little information to a time delay solution; in fact such data can
lead to ambiguous solutions. The detection of strong features in both
light curves is required to obtain a reliable time delay measurement
(\cite{kun97})\@.  Bearing this in mind, we have also analyzed the
subset of data surrounding the `bump' seen near MJD 50700 where the
most prominent and best sampled feature occurs (Figure 2(a)),
specifically to avoid ambiguities in the time delay estimate.  

We found a solution at $\Delta \tau = -26$~days, $\mu = 1.68$ and $d =
-1.17$~Jy for $\delta = 5$ (Figure 2(b)). As expected, Monte Carlo
simulations show the magnification ratio and flux density offset to be
less well constrained here than with the full dataset. We have
therefore adopted the values for $\mu$ and $d$ as determined from the
full dataset as our best estimates.  However, the simulations show the
time delay for this subset to be well constrained without ambiguities
with a Gaussian-like distribution of solutions around --26 days. By
measuring the width of this distribution at half its peak, we have
estimated an uncertainty and obtain a time delay of
$-26^{+5}_{-4}$~days.

We are continuing our ATCA monitoring observations. When the current
trend of increasing flux density at 8.6~GHz ends, the change in
gradient may provide even better constraints on the time delay.

\section{Discussion}
Out ATCA monitoring observations have allowed us to measure the time
delay and magnification ratio of the gravitational lens \et, providing
new constraints for lensing models of this system.  

Our time delay measurement is not in agreement with the $-44\pm9$ day
measurement of van Ommen \etal\ (1995) from VLA monitoring
observations at 8.6 and 15~GHz.  We believe van Ommen \etal\ did not
correctly account for the contribution of the Einstein ring flux
density when calculating the magnification ratio.  The difference
between the total flux density of the source and the sum of the
compact component flux densities fitted to the data, {\em i.e.}\ the
contribution from the ring, should be constant. However in van Ommen
\etal's analysis the inferred Einstein ring contribution depends on
the observing frequency and array configuration. We feel confident
that our analysis of the ATCA data has correctly accounted for the
constant Einstein ring flux density.

Millimeter observations, in early 1996, by Frye, Welch, \& Broadhurst
(1997), which clearly separate the core components of the lens and
show no extended jet structure on $\sim 0.5$~arcsec scales, provide an
estimate for $\mu$ of $1.14\pm0.06$\@. Also, Wiklind \& Combes (1998)
measured the magnification ratio in the millimeter band on three
occasions over 11 months and found that it changed between 1.0 and 1.8
with an uncertainty in each case of 0.1.  We note that \et\ has shown
strong variability at millimeter wavelengths in the past
(\cite{tor96}) and that Wiklind \& Combes observed a decrease in total
flux density from 1.5 to 0.93~Jy over their observations. Such
variability could influence the measurement of the magnification ratio
at a single epoch. Also, the likelihood of sub-components in the
compact structure at these wavelengths, similar to those seen at
43~GHz by Garrett \etal\ (1998), should be investigated as this too
has an impact on the magnification ratio. More detailed monitoring of
the compact components at millimeter wavelengths can provide a test of
our magnification ratio measurement if the time delay effect is
accounted for.

It is not yet possible to obtain an estimate for H$_0$ from the \et\
lensing system as the redshift of the lensed AGN is not known, and the
lensing galaxies are not well parameterized. However, if we assume
that the intervening galaxy at $z=0.19$ has a negligible gravitational
affect on the system, then we can apply our time delay measurement to
the model of Nair \etal\ (1993) for a lensing galaxy at $z=0.89$ to
put constraints on the redshift of the background source, $z_s$.  If
H$_0$ is assumed to be between 50 and 100~kms$^{-1}$Mpc$^{-1}$ and
q$_0 =1/2$, then a time delay of \delayE\ days implies $z_s > 1.4$,
with a value of $z_s \approx 2$ for H$_0 = 75$~kms$^{-1}$Mpc$^{-1}$\@.

\acknowledgments The Australia Telescope is funded by the
Commonwealth Government for operation as a national facility managed by
CSIRO\@. J.E.J.L. acknowledges the receipt of a JSPS Fellowship.

\newpage

\begin{figure}
\centerline{\rotatebox{270}{\includegraphics{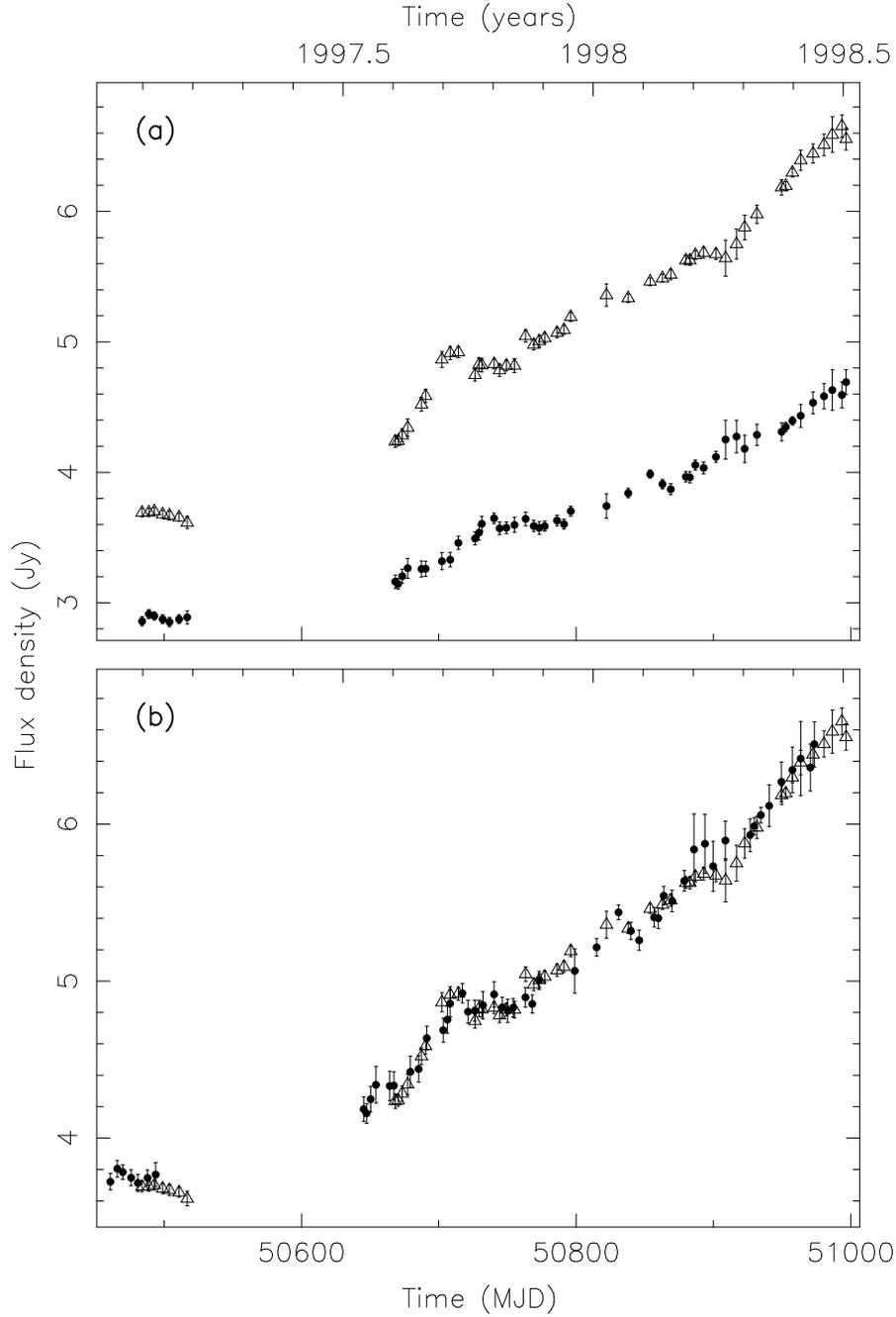}}}
\caption{(a) The $8.6$~GHz light curve data for both components of
\et. The measured NE component flux densities are 
represented by the open triangular symbols and the measured SW
component flux densities are represented by the closed circular
symbols. (b) The NE and SW component light curves after applying our
solutions from the dispersion analysis ($\Delta \tau = -23$ days, $\mu
= 1.52$ and $d = -0.62$~Jy). 
\label{fig1}}
\end{figure}

\begin{figure}
\centerline{\rotatebox{270}{\includegraphics{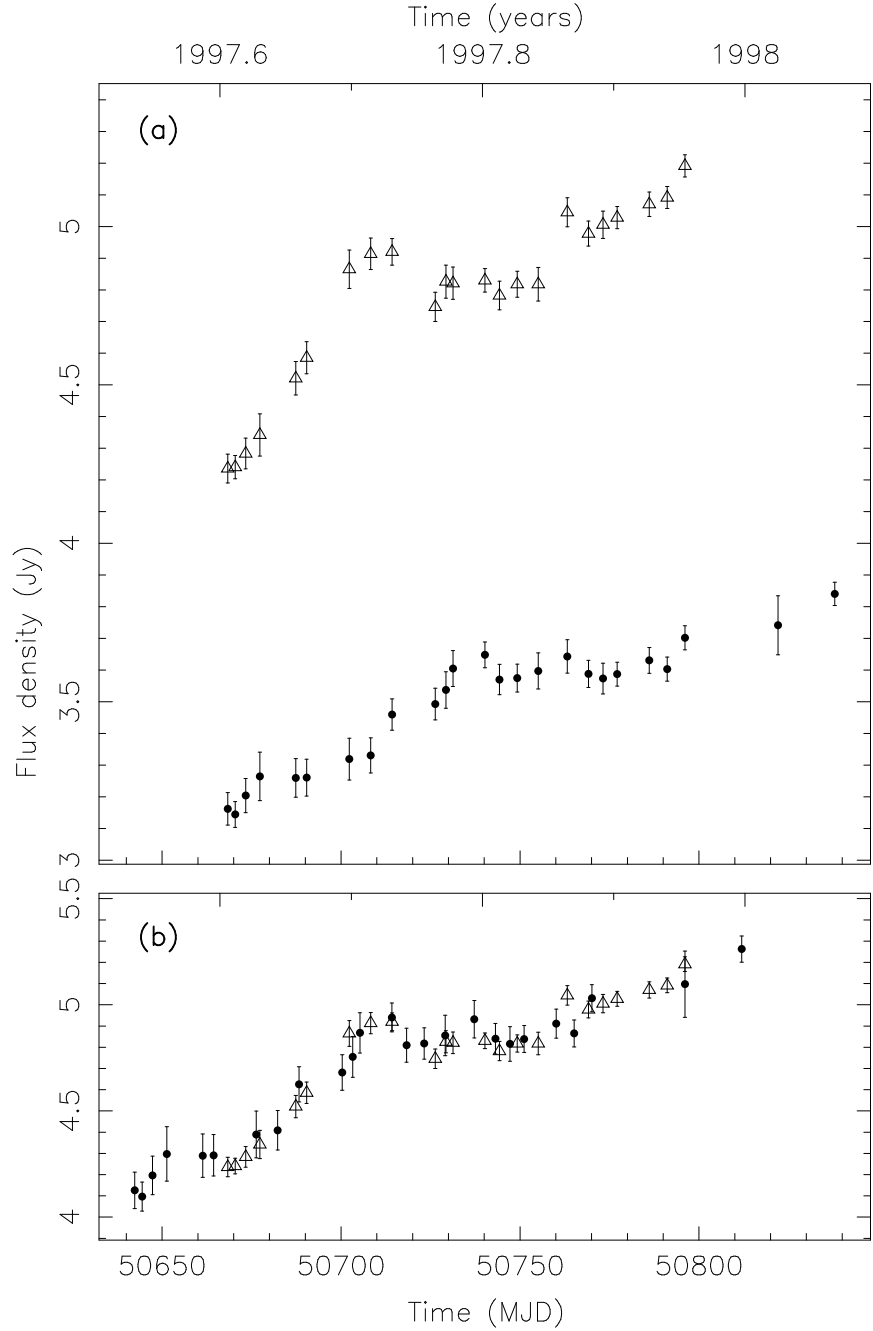}}}
\caption{(a) The $8.6$~GHz light curve data for both components of
\et\ for the time period surrounding the `bump' described in
Section~\ref{sec:obs}\@. The measured NE component flux densities are
represented by the open triangular symbols and the measured SW
component flux densities are represented by the closed circular
symbols. (b) The NE and SW component light curves after applying our
solutions from the dispersion analysis ($\Delta \tau = -26$ days, $\mu
= 1.68$ and $d = -1.17$~Jy). 
\label{fig2}}
\end{figure}
\end{document}